\documentclass[12pt]{article}
\def\beq{\begin{equation}}
\def\eeq{\end{equation}}

\begin{document}
\begin{titlepage}
\begin{center}
{\Large \bf William I. Fine Theoretical Physics Institute \\
University of Minnesota \\}  \end{center}
\vspace{0.2in}
\begin{flushright}
FTPI-MINN-04/31 \\
UMN-TH-2319-04 \\
August 2004 \\
\end{flushright}
\vspace{0.3in}
\begin{center}
{\Large \bf  Heavy quark spin selection rule and the properties of the $X(3872)$
\\}
\vspace{0.2in}
{\bf M.B. Voloshin  \\ }
William I. Fine Theoretical Physics Institute, University of
Minnesota,\\ Minneapolis, MN 55455 \\
and \\
Institute of Theoretical and Experimental Physics, Moscow, 117259
\\[0.2in]
\end{center}

\begin{abstract}
The properties of the resonance $X(3872)$ are discussed under the assumption
that this resonance is dominantly a `molecular' $J^{PC}=1^{++}$ state of neutral
$D$ and $D^*$ mesons.  It is argued that in these properties should dominate the
states with the total spin of the charmed quark-antiquark pair equal to one. As
a practical application of this observation the ratio of the rates of the decays
$X \to \pi^0 \, \chi_{cJ}$ for different $J$ is predicted. It is also pointed
out that the total rate of these decays is likely to be comparable to that of
the observed transitions $X \to \pi^+ \, \pi^- \, J\psi$ and $X \to \pi^+ \,
\pi^- \, \pi^0 \, J\psi$. The decays of the $X$ into light hadrons and its
production in hadronic processes are also briefly discussed.
\end{abstract}

\end{titlepage}

The most recent observation\cite{belle1} of the decay $X(3872) \to \pi^+ \,
\pi^- \, \pi^0 \, J/\psi$ with the rate comparable to that of the previously
observed\cite{belle2,cdf,d0,babar} process $X(3872) \to \pi^+ \, \pi^- \,
J/\psi$ unambiguously illustrates (through the obvious G parity reasoning)  that
the resonance $X(3872)$ has no definite isospin, and thus rules out an
interpretation of this resonance as a state of charmonium. Furthermore, the
spectra of the di- and tri- pion invariant masses strongly suggest that the
underlying processes in the observed decays are $X \to \omega \, J/\psi$ and $X
\to \rho^0 \, J/\psi$. The extreme proximity to the kinematical boundary in the
latter transitions favors interpreting them as $S$ wave processes, which
uniquely points to the $J^{PC}=1^{++}$ assignment for the $X(3872)$, given the
conspicuous absence of its decays into pairs of pseudoscalar $D$ mesons. Also
the fact that the mass of $X(3872)$ is within approximately $1 \, MeV$ or less
from the sum of the masses of $D^0$ and $D^{*0}$ makes it more than natural to
interpret the discussed resonance as dominantly a C-parity even S-wave near-
threshold state of the neutral meson-antimeson pair: $(D^0 {\bar D}^{*0}+ {\bar
D^0}  D^{*0})/\sqrt{2}$, much in the same way as the deuteron is dominantly a
state made of a proton and neutron. Such `molecular' systems made of charmed
mesons were suggested\cite{ov} and discussed\cite{drgg,nat,ek} long ago, and
this interpretation was considered as a favored option for the
$X(3872)$\cite{cp,ps,mv,nat2} ever since its first observation\cite{belle2}. In
particular the observed rate of the decay $X(3872) \to \pi^+ \, \pi^- \, \pi^0
\, J/\psi$ is in a very reasonable agreement with the prediction\cite{es} from a
specific model of the internal `molecular' dynamics of the $X(3872)$.

Clearly, the notion of a `molecular' state can refer only to the peripheral part
of the internal wave function of $X$, i.e at long distances beyond the range of
the strong interaction. In particular, if the mass of the $X(3872)$ is below the
$D^0 {\bar D}^{*0}$ threshold by the gap $w$, the peripheral part of its wave
function is that of a free motion of heavy mesons at the characteristic
distances set by the scale $1/\sqrt{m_D \, w}$, which is larger than about $5 \,
fm$ for $w < 1\, MeV$. At shorter distances, i.e. within the range of the strong
interaction, the mesons overlap and the ``core" of the wave function is
determined by multi-body dynamics of heavy and light quarks and gluons.
Unlike the peripheral part, which is described by just one state of the meson
pair, the properties of the core are determined by a significantly larger number
of states in the Fock decomposition
\beq
\psi_X=a_0 \, \psi_0 + \sum_i a_i \, \psi_i~,
\label{fock}
\eeq
where $\psi_0$ is the state $(D^0 {\bar D}^{*0}+ {\bar D^0}  D^{*0})/\sqrt{2}$,
while $\psi_i$ refer to `other' hadronic states. The notion of $X$ being a
`molecular' system is helpful inasmuch as the probability weight $|a_0|^2$ of
the meson component $\psi_0$ makes a large portion of the total normalization.
In particular the model of Ref.\cite{es} includes $S$ and $D$ wave states of the
neutral and charged charmed meson pairs as well as the channels $\rho J/\psi$
and $\omega J/\psi$, and estimates the weight factor of the $S$ wave $(D^0 {\bar
D}^{*0}+ {\bar D^0}  D^{*0})$ component as 70-80\% at $w \approx 1 \, MeV$.

It can be noted however, that although the peripheral `molecular' component
$\psi_0$ has the largest probability, only a limited scope of properties of the
$X$ can be understood using only this component, such as (yet unobserved) decays
$X \to D^0 \, {\bar D}^0 \, \pi^0$ and  $X \to D^0 \, {\bar D}^0 \,
\gamma$\cite{mv}, for which the underlying processes is the decay of individual
$D^*$ (${\bar D}^*$) meson. The majority of the production and decay properties
of the $X$ are determined by distances at least as short as the confinement
range and are sensitive to the dynamics in the core. These properties include in
particular the production of the $X(3872)$ in $B$ decays and in hadronic
collisions, as well as the decays of $X$ into final states with charmonium
resonances, and into light hadrons. Given the present state of understanding the
strong dynamics in the confinement region, it appears that any guidelines from
the general properties of QCD, even quite approximate, may serve as a useful
starting point in the studies of these properties.

The purpose of this paper is to point out a simple approximate spin selection
rule for the charmed quark-antiquark pair in $X(3872)$ considered as a
$J^{PC}=1^{++}$ `molecular' system in the previously described sense.
The parameter for applicability of this rule is $\Lambda_{QCD}/m_c$. Although
for the realistic charmed quark mass this parameter does not look reliably
small, it still might be helpful for understanding the processes involving
$X(3872)$. Namely, it will be argued here that up to corrections of order
$\Lambda_{QCD}/m_c$, the Fock sum in eq.(\ref{fock}) should contain only the
states where the total spin $S_{c {\bar c}}$ of the $c {\bar c}$ quark pair is
equal to one: $S_{c {\bar c}}=1$, irrespectively of the angular momentum and/or
the overall color of this pair. Conversely, the coefficients of the states with
$S_{c {\bar c}}=0$ in the Fock sum should be suppressed by the ``small"
parameter\footnote{The coefficients of the states without the hidden charm, i.e.
only with light quarks/hadrons are trivially suppressed by $m_c^{-1}$, since the
charmed quarks have to annihilate at distances $\sim m_c^{-1}$.}.

As an illustration the decays $X \to \pi^0 \, \chi_{cJ}$ will be discussed here,
for which the spin selection rule determines the relative rate of the
transitions to the charmonium states with different $J$.

The argumentation for the discussed here selection rule can be started with
noticing that in the $S$-wave C-even meson pair $(D^0 {\bar D}^{*0}+ {\bar D^0}
D^{*0})$ the $c {\bar c}$ quark pair is necessarily in the pure spin state with
$S_{c {\bar c}}=1$. Indeed, if such meson pair is considered as a four-quark
state $c {\bar c} \, u {\bar u}$ with all the relative orbital angular momenta
equal to zero, the total angular momentum $J=1$ is determined as the vector sum
of the spins of the quarks and antiquarks. Each of the quark pairs, $c {\bar c}$
and $u {\bar u}$ generally can have either $S=1$ or $S=0$. However, one can
readily see that the states where $J=1$ arises from combining $S=1$ from one
pair with $S=0$ of the other pair have negative C parity. The only combination
resulting in a C-even $J=1$ state is where each quark-antiquark pair has $S=1$.
Naturally, the total spin of the light quark pair is not ``traceable" since it
is changed by the strong interaction with amplitude of order one. Indeed, the
energy gap between the states with different $S_{u {\bar u}}$ is of order
$\Lambda_{QCD}$ and the strong mixing amplitude is of the same order. The
situation however is quite different for the total spin of the heavy quark pair.
The spin-flip amplitude for a heavy charmed quark contains the factor $m_c^{-
1}$, while the energy gap is still of order $\Lambda_{QCD}$. It should be
emphasized that the latter gap arises not from the spin-dependent interactions
within the $c {\bar c}$ pair (which interactions have another extra factor of
$m_c^{-1}$), but rather from the rearrangement of the spin and/or orbital state
of the light quark pair and/or changing the number of `valence' gluons in the
wave function, imposed by the conservation of the total angular momentum of the
system and its P and C parities. Thus the sates $\psi_i$ in the sum in
eq.(\ref{fock}), resulting from the strong-interaction mixing with the
`molecular' meson state, should obey the stated spin selection rule.

Applying the spin selection rule to hadronic transitions from the $X(3872)$ to
charmonium levels, one readily concludes that such transitions to the
spin-singlet para-charmonium levels, e.g. $X \to \pi \, \pi \, \eta_c$ should be
suppressed, while those to the spin-triplet ones should be favored. The latter
include the observed decays $X(3872) \to \pi^+ \, \pi^- \, J/\psi$ and $X(3872)
\to \pi^+ \, \pi^- \, \pi^0 \, J/\psi$ and also the yet unobserved (and
apparently not yet mentioned in the literature) decays to the spin-triplet
$P$-wave states $\chi_{cJ}$. The kinematics and the quantum numbers allow such
decays with a $P$-wave emission of a single $\pi^0$: $X \to \pi^0 \, \chi_{cJ}$.
The discussed here conservation of the total spin of the heavy quark-antiquark
pair implies the relation between the rates of these decays into states with
different $J$:
\beq
\Gamma(X \to \pi^0 \, \chi_{cJ}) \propto (2J+1)\, p_\pi^3~,
\label{gj}
\eeq
where $p_\pi$ is the momentum of the pion.

Understandably, there is a great uncertainty in estimating the absolute rate of
the decays $X \to \pi^0 \, \chi_{cJ}$. It is clear that as compared to the decay
$X \to \pi^+ \, \pi^- \, J\psi$, viewed as $X \to \rho J/\psi$, the amplitude of
these decays should contain a dimensional factor $\mu^{-1}$ describing the
excitation of the heavy quark $P$-wave, which should naturally be of order of
the characteristic size of the component of the core with the $c {\bar c}$ pair
being in $P$ wave. In terms of this factor the ratio of the decay rates can be
estimated as
\beq
{\Gamma(X \to \pi^0 \, \chi_{cJ}) \over \Gamma(X \to \pi^+ \, \pi^- \, J\psi)}=
{2J +1 \over 9}\, {p_\pi^3 \over   (q_\rho)_{eff} \, \mu^2}~,
\label{rat}
\eeq
where $(q_\rho)_{eff}$ is the effective momentum of the $\rho$ meson in the
decay $X \to \pi^+ \, \pi^- \, J\psi$ reflecting the fact that the $\rho^0$
materializes in the process as two pions rather than as a single particle. The
value of  $(q_\rho)_{eff}$ is found as
\beq
(q_\rho)_{eff}=\int_{4m_\pi^2}^{\Delta^2} \sqrt{\Delta^2-q^2} \, {m_\rho \,
\Gamma_\rho(q^2) \over (q^2-m_\rho^2)^2+m_\rho^2 \, \Gamma_\rho^2(q^2)} \, {dq^2
\over \pi} \approx 120 \, MeV~,
\label{qeff}
\eeq
where $\Delta=M(X)-m_\rho$, and $\Gamma_\rho(q^2)$ is the width parameter of the
$\rho$ meson with $\Gamma_\rho(m_\rho^2)=\Gamma_\rho$. Using then eq.(\ref{rat})
for an estimate of the decay with e.g. the $\chi_{c1}$ in the final state (the
most advantageous from the point of tagging through $\chi_{c1} \to \gamma \,
J/\psi$), one finds
\beq
{\Gamma(X \to \pi^0 \, \chi_{c1}) \over \Gamma(X \to \pi^+ \, \pi^- \, J\psi)}
\approx 0.35 \left ( {0.5 \,  GeV \over \mu} \right )^2~,
\label{est}
\eeq
which shows that the discussed decays should have a realistically observable
rate for reasonable values of the parameter $\mu$.

Considering the decays of the $X(3872)$ into light hadrons, where the $c {\bar
c}$ quark pair has to annihilate, one can also use the suggested spin selection
rule in combination with the ``charm burning" mechanism\cite{ov}, according to
which the charmed quarks can annihilate from a `molecular' state being not
necessarily in a color-singlet state of the $c {\bar c}$ pair. The spin
selection rule however requires the annihilation to proceed from a spin-triplet
state. Then the annihilation rate is the largest from the color-octet $^3S_1$
state and is determined by the annihilation $c {\bar c} \to q {\bar q}$ through
one gluon, since such state does not annihilate into two gluons in the lowest
order in $\alpha_s$\cite{bgr,ov2}. The rate of such decay is determined by the
relevant size parameter of the core $(\mu^{\prime})^{-1}$ and by the probability
weight factor of the core:
\beq
\Gamma(X \to {\rm light~hadrons}) \sim |a_{core}|^2 \alpha_s^2(m_c)
{(\mu^{\prime})^3 \over m_c^2}~.
\label{est2}
\eeq
Given that the core probability weight is perceived as ``few percent" and that
the $O(\alpha_s^2)$ annihilation rate of the charmonium states is ``few MeV",
the best estimate of the factors in this relation can at present be formulated
only as ``few percent of few MeV", i.e. in the range from about a hundred to few
hundred KeV. As uncertain as such guesstimate of the annihilation rate is, it is
at least not in an apparent disagreement with the known experimental facts about
the $X(3872)$.

The discussed spin selection rule might also be helpful in understanding the
production of the $X(3872)$ in hadronic processes. Indeed, the production of
this resonance through its peripheral `molecular' component, i.e. by coalescence
of the charmed mesons is extremely weak\cite{bk}, so that the actual processes
proceeds through production of the core component of the $X$. It can be remarked
that  (irrespectively of the discussed spin selection rule) this picture agrees,
at least semi-quantitatively, with the rate\cite{belle2} of the observed decays
of $B \to K \, X(3872)$:
\beq
{{\cal B}(B^+ \to K^+ \, X) \, {\cal B}(X \to \pi^+ \, \pi^- \, J/\psi) \over
{\cal B}(B^+ \to K^+ \, \psi^{\prime}) \, {\cal B}(\psi^{\prime} \to \pi^+ \,
\pi^- \, J/\psi)} = 0.063 \pm 0.014~.
\label{dat}
\eeq
Indeed, it is an experimental fact that the known charmonium states are produced
in $B$ decays in association with a single Kaon with approximately the same rate
(within a factor of two). One might expect then that the core states of the
$X(3872)$ are produced in similar decays at approximately the same rate, so that
the only suppression factor is that of the probability weight $|a_{core}|^2$,
i.e. few percent. Thus there is no dramatic disagreement with the experimental
number (\ref{dat}), if one reasonably assumes  that  the branching ratio ${\cal
B}(X \to \pi^+ \, \pi^- \, J/\psi)$ is not too small in comparison with ${\cal
B}(\psi^{\prime} \to \pi^+ \, \pi^- \, J/\psi) \approx 0.3$.

The spin selection rule obviously somewhat restricts the possible states of the
$c {\bar c}$ pair produced in a hadronic process, which fragment into the core
of the $X(3872)$. It should be noted that this rule is less restrictive than
recently suggested by Braaten\cite{eb}, where only the production of a
color-singlet $^3P_1$ and a color-octet $^3S_1$ $c {\bar c}$ states are
considered (with further assumptions about the relative contributions of these
mechanisms). The discussed here spin selection rule generally allows production
of the $X(3872)$ originating from other states of $c {\bar c}$ as long as those
states are spin triplets. It is a matter of further study whether the more
restrictive assumptions of Ref.\cite{eb} are applicable in the actual production
processes.

In summary. The interpretation of the resonance $X(3872)$ as a $J^{PC}=1^{++}$
`molecular' state of neutral $D$ and $D^*$ mesons is gaining support from
experimental data. It is argued here that among the configurations present in
the wave function describing the structure of such state in terms of quarks and
gluons, should dominate those where the total spin of the $c {\bar c}$ pair is
equal to one: $S_{c {\bar c}}=1$. The parameter for suppression of the
spin-singlet configurations is $\Lambda_{QCD}/m_c$. In particular this rule
predicts suppression of decays of the $X$ resonance into spin-singlet charmonium
states, such as e.g. $X \to \pi \, \pi \, \eta_c$, in comparison with its
hadronic transitions to spin-triplet charmonium. The latter transitions include
the observed decays with the $J/\psi$ resonance in the final state and
also the yet unobserved decays $X \to \pi^0 \, \chi_{cJ}$, for which the spin
selection rule determines the ratio of the rates, and their absolute rate is
likely to be within the reach of experiment. The spin selection rule favors the
pattern of the decay of the $X$ resonance into light hadrons determined by the
annihilation of the $c {\bar c}$ pair into light quarks through one gluon: $c
{\bar c} \to q {\bar q}$, and also the same rule can be helpful in the studies
of the production of $X(3872)$ in hadronic processes.

This work is supported in part by the DOE grant DE-FG02-94ER40823.

\end{document}